\documentclass[journal=jacsat,manuscript=communication,etalmode=truncate,keywords=true]{achemso}
\setkeys{acs}{articletitle=true,maxauthors=10,etalmode=truncate,keywords=true}
\usepackage{amsmath,amssymb,units,txfonts}
\usepackage{amsfonts}
\usepackage{graphicx,MnSymbol}
\usepackage{helvet,wasysym}
\usepackage[bookmarks=true,colorlinks=true,urlcolor=blue,linkcolor=blue,citecolor=blue]{hyperref} 
\usepackage[usenames,dvipsnames]{color}
\usepackage{colortbl,cuted}

\definecolor{JPCCBlue}{RGB}{34,80,169}
\definecolor{NLRed}{RGB}{215,24,30}
\definecolor{abstractcolor}{RGB}{255,243,201}
\makeatletter\newenvironment{abstractbox}{%
   \begin{lrbox}{\@tempboxa}\begin{minipage}{0.988\textwidth}}{\end{minipage}\end{lrbox}%
   \colorbox{abstractcolor}{\usebox{\@tempboxa}}
}\makeatother

\renewcommand*\authorfont{\flushleft}
\renewcommand*\affilfont{\mdseries\flushleft\textrm}
\renewcommand*\titlesize{\Large}
\renewcommand*\titlefont{\flushleft\sf\bfseries}
\def\refname{\sffamily\color{JPCCBlue}\Large$\blacksquare$\normalsize\ REFERENCES}
\usepackage{titlesec}
\titleformat{\section}{\bfseries\sffamily\color{JPCCBlue}}{\thesection.~}{0pt}{}
\titleformat{\subsection}[runin]{\bfseries\sffamily\normalsize}{\indent\thesubsection.~}{0pt}{}[.]
\titlespacing{\subsection}{0pt}{0pt}{*1}
\titleformat{\subsubsection}{\bfseries\sffamily\normalsize}{\thethesubsection.~}{0pt}{}
\titlespacing{\subsubsection}{0pt}{0pt}{*0}

\title{Gold and Methane: A Noble Combination for Delicate Oxidation}

\author{Duncan J.~Mowbray}
\email{duncan.mowbray@gmail.com}
\affiliation[UPV/EHU]{\footnotemark[2]{\ } Nano-Bio Spectroscopy Group and ETSF Scientific Development Center, Departamento de F{\'{\i}}sica de Materiales, Centro de F\'{\i}sica de Materiales CSIC-UPV/EHU-MPC and DIPC, Universidad del Pa{\'{\i}}s Vasco UPV/EHU, E-20018 San Sebasti\'{a}n, Spain}
\author{Annapaola Migani}
\email{annapaola.migani@cin2.es}
\affiliation[UPV/EHU]{\footnotemark[2]{\ } Nano-Bio Spectroscopy Group and ETSF Scientific Development Center, Departamento de F{\'{\i}}sica de Materiales, Centro de F\'{\i}sica de Materiales CSIC-UPV/EHU-MPC and DIPC, Universidad del Pa{\'{\i}}s Vasco UPV/EHU, E-20018 San Sebasti\'{a}n, Spain}
\alsoaffiliation[CIN2]{\newline\footnotemark[3]{\ } CSIC - Consejo Superior de Investigaciones Cientificas, ICN2 Building, E-08193 Bellaterra (Barcelona), Spain}
\author{Guido Walther}
\affiliation[LIKAT]{\newline\footnotemark[5]{\ } Leibniz Institute for Catalysis at the University of Rostock, D-18059 Rostock, Germany}
\author{David M.~Cardamone}
\author{Angel Rubio}
\email{angel.rubio@ehu.es}
\affiliation[UPV/EHU]{\footnotemark[2]{\ } Nano-Bio Spectroscopy Group and ETSF Scientific Development Center, Departamento de F{\'{\i}}sica de Materiales, Centro de F\'{\i}sica de Materiales CSIC-UPV/EHU-MPC and DIPC, Universidad del Pa{\'{\i}}s Vasco UPV/EHU, E-20018 San Sebasti\'{a}n, Spain}

\begin{document}
\maketitle

\begin{strip}
\vspace{-1.cm}

\noindent{\color{JPCCBlue}{\rule{\textwidth}{0.5pt}}}
\begin{abstractbox}
\begin{tabular*}{17cm}{b{11.6cm}r}
\noindent\textbf{\color{JPCCBlue}{ABSTRACT:}}
The ability to partially oxidize methane at low temperatures and pressures would have important environmental and economic applications. Although methane oxidation on gold nanoparticles has been observed experimentally, our density functional theory (DFT) calculations indicate neither CH$_{\text{4}}$, CH$_{\text{3}}$, nor H adsorb on a \emph{neutral} gold nanoparticle. However, by positively \emph{charging} gold nanoparticles, e.g.\ through charge transfer to the TiO$_{\text{2}}$ substrate, CH$_{\text{4}}$ binding increases while O$_{\text{2}}$ binding remains relatively unchanged. We demonstrate that CH$_{\text{4}}$ adsorption is via bonding with the metal \emph{s} levels.  This holds from small gold clusters (Au$_{\text{2}}$) to large gold nanoparticles (Au$_{\text{201}}$), and for all fcc transition metal dimers.  These results provide the chemical understanding necessary to tune the catalytic activity of metal nanoparticles for the partial oxidation of methane under delicate conditions.
\newline
\newline
{{{\color{JPCCBlue}{\textbf{KEYWORDS:}}} {CH$_{\text{4}}$ $\cdot$ Au$_n$ $\cdot$ nanoparticles $\cdot$ O$_{\text{2}}$ $\cdot$ DFT $\cdot$ heterogeneous catalysis}}}

{\textbf{\color{JPCCBlue}{SECTION:}}  Surfaces, Interfaces, Porous Materials, and Catalysis}
&\includegraphics[height=5cm]{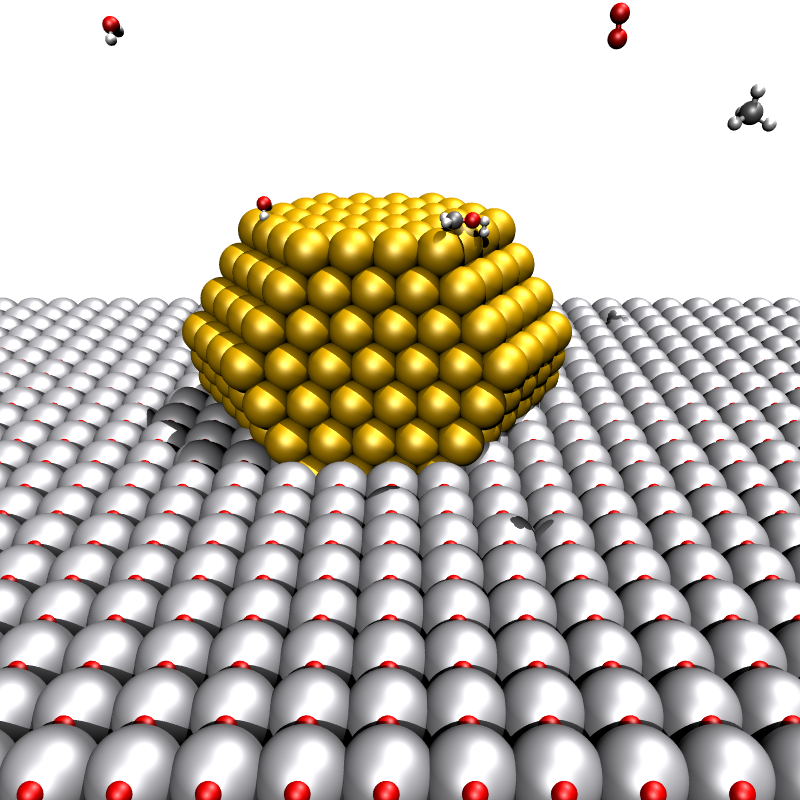}\\
\end{tabular*}
\end{abstractbox}
\noindent{\color{JPCCBlue}{\rule{\textwidth}{0.5pt}}}
\end{strip}

\def\bigfirstletter#1#2{{\noindent
    \setbox0\hbox{{\color{JPCCBlue}{\Huge #1}}}\setbox1\hbox{#2}\setbox2\hbox{(}%
    \count0=\ht0\advance\count0 by\dp0\count1\baselineskip
    \advance\count0 by-\ht1\advance\count0 by\ht2
    \dimen1=.5ex\advance\count0 by\dimen1\divide\count0 by\count1
    \advance\count0 by1\dimen0\wd0
    \advance\dimen0 by.25em\dimen1=\ht0\advance\dimen1 by-\ht1
    \global\hangindent\dimen0\global\hangafter-\count0
    \hskip-\dimen0\setbox0\hbox to\dimen0{\raise-\dimen1\box0\hss}%
    \dp0=0in\ht0=0in\box0}#2}
\bigfirstletter
Understanding how one can control and manipulate the catalytic activity of nanoparticles is a major current challenge for nanoscience. One of the most important applications of such a technology, with significant environmental implications, is the partial oxidation of methane \cite{Concepts,MethaneCatalysis,HutchingsMethane} under mild conditions by gold nanoparticles \cite{Haruta1993175,HarutaGoldBulletin,HarutaNature,CatalysisbyGold,GoldReview,NorskovNanoToday,Mystery}.

In this letter, we explore the dependence of methane adsorption on gold cluster size, from gold dimers to large gold nanoparticles. We use density functional theory (DFT) to develop theoretical models, which in turn provide chemical insight into recent experimental results for methane oxidation on gold nanoparticles \cite{GuidoCatToday,GuidoGoldBulletin}.  By positively charging gold nanoparticles, we show one can tune the alignment of the \emph{s} levels to control methane adsorption independently from O$_2$ adsorption, and hence their catalytic activity.

These results suggest a direct catalytic pathway from a naturally abundant greenhouse gas to more valuable and useful chemicals.  Specifically, by tuning the adsorption energy of CH$_{\text{4}}$ independently of O$_{\text{2}}$ via charging, controlled partial oxidation of methane under delicate conditions can be achieved at the nanoscale.
This is because the adsorption energies of reactants typically determine the activity of reactions in heterogeneous catalysis \cite{HydrogenEvolution}.  In general, reactants must adsorb neither so strongly they are unable to migrate and react, nor so weakly they are unable to react prior to desorbing \cite{MowbrayJPCC}. Therefore, an essential first step in controlling methane oxidation is to understand and control the adsorption properties of methane and oxygen on gold nanoparticles.

The adsorption and dissociation of oxygen on gold clusters and nanoparticles has already received much attention in the literature \cite{MowbrayJPCC,LandmanO2,Metiu2008,Illas,AskPRB}.  These studies have shown that charging, size, and atomic coordination are the main factors which control the adsorption and dissociation energies of oxygen on gold.  

However, the mechanism of interaction between methane and gold has received much less attention.  Studies have focused on either Au$^{\text{I}}$ and Au$^{\text{III}}$ oxidized species \cite{Goddard} or small gold cluster cations Au$_n^+$, $n=$2--6 \cite{LandmanAngewandte,LandmanCPC,LandmanCH4Au2,BernhardtFaradayDiscussions}.  The common characteristic of these systems is that Au is positively charged.  This suggests that the gold---methane interaction might be charge-transfer mediated.  

\renewcommand{\thefootnote}{}
\footnotetext{\hspace{-0.5cm}This document is the unedited Author's version of a Submitted Work that was subsequently accepted for publication in The Journal of Physical Chemistry Letters, copyright \copyright American Chemical Society after peer review. To access the final edited and published work see \href{http://dx.doi.org/10.1021/jz401553p}{http://dx.doi.org/10.1021/jz401553p}.}

In heterogeneous catalysis, methane oxidation on metal oxide supported gold nanoparticles of diameter 1.8 nm has recently been demonstrated experimentally\cite{GuidoCatToday,GuidoGoldBulletin}.   However, oxidation only occurred at higher temperatures, and was always complete \cite{GuidoCatToday,GuidoGoldBulletin}.  No oxidation occured on the clean metal oxide support.\cite{GuidoThesis}

These results raise the following questions. How does methane adsorb on gold nanoparticles? Is it due to charging of the nanoparticle, perhaps coming from the metal oxide support? \cite{Metiu2007,Metiu2007I,Metiu2007II}  Can the charging of the nanoparticle be used as a descriptor to predict and control the adsorption of methane on gold?

Guided by these observations, we have undertaken a systematic study within DFT of CH$_{\text{4}}$ and O$_{\text{2}}$ adsorption on gold clusters and nanoparticles.  We find a linear correlation between the adsorption energy for CH$_{\text{4}}$ and the charge transferred from methane to gold. This relation holds independently of the size of the gold species (Au$_n$ for $n=$ 2, 6, 7, 55, 201) and the coordination number of the adsorption site ($N_c=$ 1, 2, 3, 5, 6). 
Moreover, it holds for gold species in vacuum 
and on a pristine or defective rutile TiO$_{\text{2}}$(110) support.  

\begin{figure}[!tb]
\includegraphics[width=0.99\columnwidth]{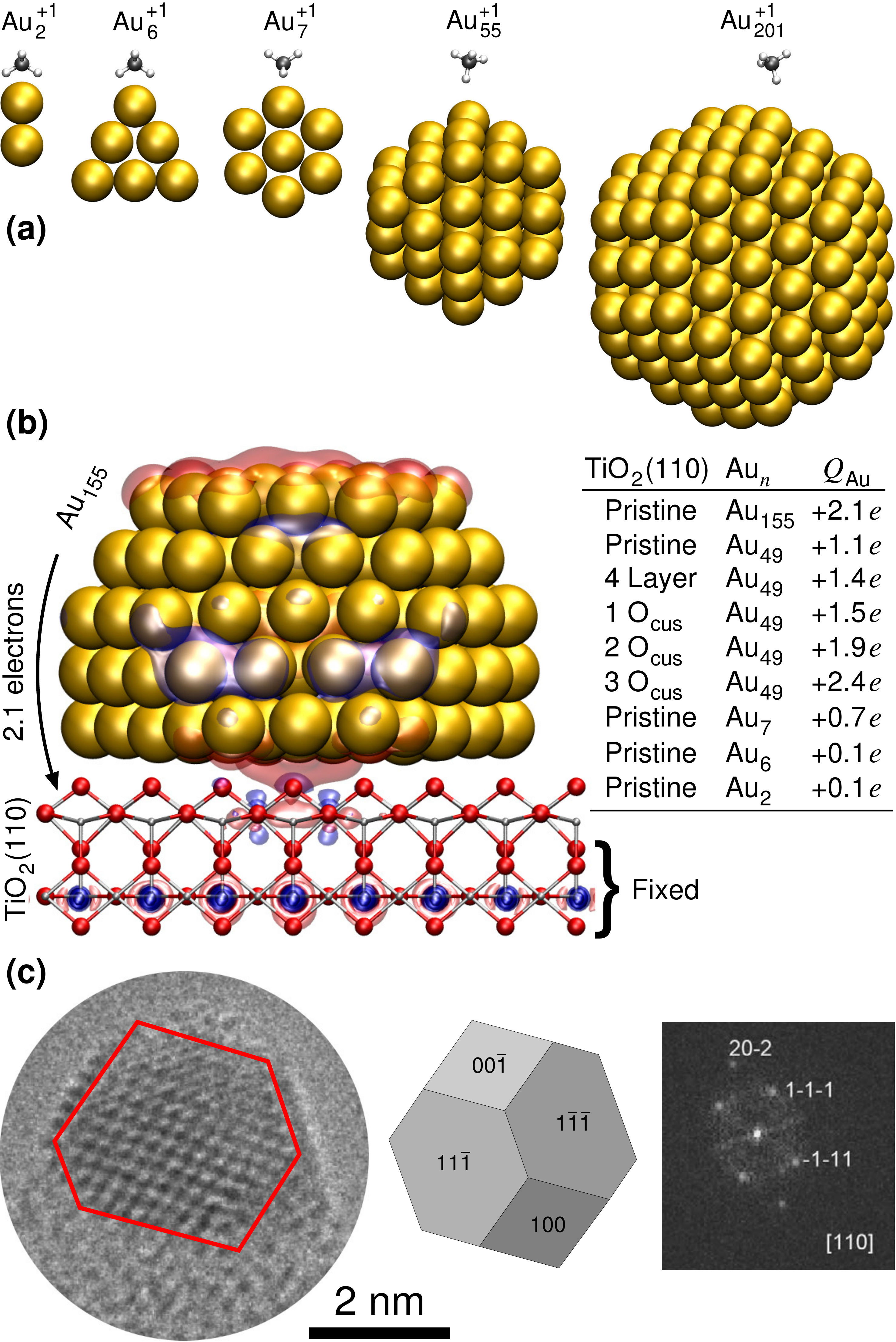}
\caption{(a) Schematics of CH$_{\text{4}}$ adsorption on Au$_{\text{2}}^{\text{+1}}$, Au$_{\text{6}}^{\text{+1}}$, Au$_{\text{7}}^{\text{+1}}$, Au$_{\text{55}}^{\text{+1}}$, and Au$_{\text{201}}^{\text{+1}}$.  
(b)  Schematic of Au$_{\text{155}}$ adsorbed on a pristine rutile TiO$_{\text{2}}$(110) surface, and charge $Q_{\text{Au}}$ in $e$ of the adsorbed Au$_n$ clusters on pristine or defective rutile TiO$_{\text{2}}$(110).  
(c) High resolution TEM image of a 2.8 nm gold particle supported on rutile TiO$_{\text{2}}$ after methane oxidation.  The particle is cuboctahedral, as emphasized by the red border, and adheres to the support with its \{111\} surface. A 3D reconstruction and diffraction pattern are shown to the right \cite{GuidoThesis}.
}\label{fgr:Fig1} 
\noindent{\color{JPCCBlue}{\rule{\columnwidth}{1pt}}}
\end{figure}

The systems we consider for methane adsorption span the range from 
the cationic species used in gas phase experiments \cite{LandmanCH4Au2,BernhardtFaradayDiscussions} (Au$_{\text{2}}^{\text{+1}}$) to the truncated cuboctahedral nanoparticle utilized in heterogeneous catalysis experiments \cite{GuidoCatToday,GuidoGoldBulletin} (Au$_{\text{201}}$) shown in Figure~\ref{fgr:Fig1}(a).  
A linear correlation between methane adsorption and charging holds from 
the smallest cluster to a large nanoparticle, both in gas phase and including the support effect.  For this reason, we propose this correlation as a unifying concept for understanding the interaction between gold and noble species in gas phase and heterogeneous catalysis experiments.  Moreover, based on the alignment of the gold 6\emph{s} states with the methane 2$a_{\text{1}}$ level, we formulate two semi-quantitative models to predict methane adsorption energies on charged gold nanoparticles.

On the other hand, the O$_{\text{2}}$ adsorption energy has a much weaker dependence on the charge transfer for positively charged gold nanoparticles. Rather, as previously shown, we find a strong dependence on the gold particle size and the coordination number of the adsorption site \cite{MowbrayJPCC}.  This suggests one may tune the adsorption of CH$_{\text{4}}$ through charging without changing the adsorption of O$_{\text{2}}$.  We shall show that this has important consequences for performing delicate oxidation of CH$_{\text{4}}$ on gold nanoparticles.

All DFT calculations were performed within the Grid-based Projector-Augmented Wavefunction method code GPAW \cite{gpaw1,gpaw2}. A grid spacing of 0.2~\AA\ was used, with all structures relaxed until a maximum force of less than 0.05 eV/\AA\ was obtained, as provided in Supporting Information.  We employ the RPBE exchange-correlation (xc)-functional. \cite{RPBE} RPBE has been shown to provide a similar degree of accuracy to the more computationally expensive B3LYP hybrid functional for adsorption properties on metal surfaces \cite{RPBEvsB3LYP}. A comparison between CH$_4$ adsorption eneriges on Au$_2$ for RPBE, PBE\cite{PBE}, and vdW-DF\cite{vdW-DF} xc-functionals is provided in Supporting Information. Non-periodic boundary conditions have been applied for all charged calculations, employing more than 5~\AA\ of vacuum to the cell boundary, where both the electron density and wavefunctions are set to zero.  This corresponds to the vacuum level $E_{\textit{vac}}$ in our calculations. This procedure removes spurious electrostatic interactions, which would be quite strong in periodic calculations for charged systems.  

A molecule's adsorption energy $E_{\textit{ads}}$ is defined as the difference in total energy between the molecule adsorbed on the gold species and the separated molecule in gas phase and clean gold species.  For example, for CH$_{\text{4}}$ on an $n$ atom gold nanoparticle of charge $Q$, Au$_n^Q$, the adsorption energy is 
\begin{equation}
E_{\textit{ads}}[\text{CH}_{\text{4}}] = E[\text{CH}_{\text{4}}-\text{Au}_n^Q] - E[\text{CH}_{\text{4}}] - E[\text{Au}_n^Q].
\end{equation}
Note that the gas phase energy for O$_{\text{2}}$ is obtained from the H$_{\text{2}}$O formation reaction, as described in Ref.~\citenum{MowbrayO2Ref}\nocite{MowbrayO2Ref} and the Supporting Information.  

There are two commonly used models for the catalytic activity of gold nanoparticles.  The model of Haruta \emph{et al.} \cite{Haruta1993175,HarutaGoldBulletin,HarutaNature,CatalysisbyGold} attributes the catalytic activity of gold nanoparticles to the perimeter of the gold--support interface.  Since the reaction rate $R$ scales as the number of active sites per unit volume, in this case $R \sim \nicefrac{1}{d^{2}}$, where $d$ is the nanoparticle's diameter.  On the other hand, the model of N{\o}rskov \emph{et al.} \cite{NorskovNanoToday,NorskovJCat,FalsigCO,MowbrayJPCC} attributes the catalytic activity to the most undercoordinated sites of the gold nanoparticles, i.e.\ the corners. In this case the reaction rate scales inversely with the volume so that $R \sim \nicefrac{1}{d^{3}}$.  For CO oxidation on gold nanoparticles with $d \lesssim 2$~\AA, the experimentally measured reaction rate scales as $\nicefrac{1}{d^{3}}$ independently of the support used  \cite{NorskovNanoToday}.  Thus CO oxidation occurs primarily at the corners of the gold nanoparticles in this size regime. 
However, for larger nanoparticles ($d \gtrsim 2$~\AA) the perimeter of the gold--support interface begins to play a role, as seen in the scaling of the measured reaction rate for CO oxidation \cite{NorskovNanoToday}.

The apparent activation barrier for CO oxidation on gold nanoparticles was quantitatively well described by DFT-based microkinetic models that only consider the corner sites \cite{Guido}. These results revealed the different reaction mechanisms responsible for CO oxidation on gold nanoparticles when O$_{\text{2}}$ or N$_{\text{2}}$O is used as the oxidant \cite{Guido}.
Since here our primary interest is to model methane oxidation experiments performed on gold nanoparticles with $d \lesssim 2$~\AA, we shall focus on CH$_4$ adsorption directly on the gold nanoparticle. 
To this end, we have performed DFT calculations on a neutral Au$_{\text{55}}$ cluster in vacuum to determine the most strongly binding sites.

Methane behaves as a noble gas. This means adsorption of methane on metals is usually weak. For this reason, previous studies on other metals assumed methane adsorption is dissociative (i.e.\ as CH$_{\text{3}}$ and H) and not the rate-determining step\cite{NorskovMethaneDecomposition}. However, for the noble combination of gold and methane, we find the coadsorption energy for CH$_{\text{3}}$ and H on Au$_{\text{55}}$ (1.18~eV) is already more than twice the measured apparent activation barrier for methane oxidation on gold nanoparticles (0.52~eV).\cite{GuidoCatToday,GuidoGoldBulletin}  Therefore, methane does not adsorb on gold nanoparticles dissociatively.  Our calculations also show that neither O$_{\text{2}}$-assisted dissociative adsorption nor O-assisted adsorption is the mechanism of methane oxidation on gold nanoparticles.

To determine whether the adsorption of methane on gold nanoparticles may be mediated by charge transfer to the substrate, we calculated the charge $Q_{\text{Au}}$ of Au$_n$ upon adsorption on a pristine or defective rutile TiO$_{\text{2}}$(110) surface. The charge transfer is obtained from a Bader analysis\cite{bader} of the all-electron charge density.  We model the adsorbed gold nanoparticles using Au$_{\text{155}}$ and Au$_{\text{49}}$, corresponding to the top two-thirds of pristine Au$_{\text{201}}$ and Au$_{\text{55}}$, on a frozen two or four titanium layer rutile TiO$_{\text{2}}$(110) surface, as illustrated in Figure \ref{fgr:Fig1}(b).  This is done to better approximate the experimentally observed geometry of gold nanoparticles on TiO$_{\text{2}}$ after methane oxidation, shown in Figure~\ref{fgr:Fig1}(c).  We model surface defects by displacing a bridging O atom far from the cluster to a coordinately unsaturated Ti site (cus) below the nanoparticle, O$_{\text{cus}}$.  In this way we ensure the system remains stoichiometric, while including typical surface defects observed on TiO$_{\text{2}}$(110) \cite{DieboldTiO2,TiO2110Defects}. 

As shown in Figure~\ref{fgr:Fig1}(b), all the Au$_n$ species considered become positively charged upon adsorption on pristine TiO$_{\text{2}}$(110), with the charge transfer increasing with the particle size.  This agrees with previous studies for Au$_{\text{2}}$, Au$_{\text{6}}$, and Au$_{\text{7}}$ \cite{Metiu2007,Metiu2007I,Metiu2007II}.  For Au$_{\text{49}}$ and Au$_{\text{155}}$ we find the observed charge transfer is from the Au 6\emph{s} levels to the Ti 3\emph{d} levels, as illustrated in Figure~\ref{fgr:Fig1}(b). These Ti 3\emph{d} levels constitute the conduction band of TiO$_{\text{2}}$.  By increasing the number of TiO$_{\text{2}}$ layers or O$_{\text{cus}}$ atoms, we systematically increase the positive charge on Au$_{\text{49}}$.  Altogether, this suggests positively charged gold nanoparticles could play an important role in methane oxidation experiments.  

\begin{figure}
\noindent{\color{JPCCBlue}{\rule{\columnwidth}{1pt}}}
\begin{center}
\includegraphics[width=0.97\columnwidth]{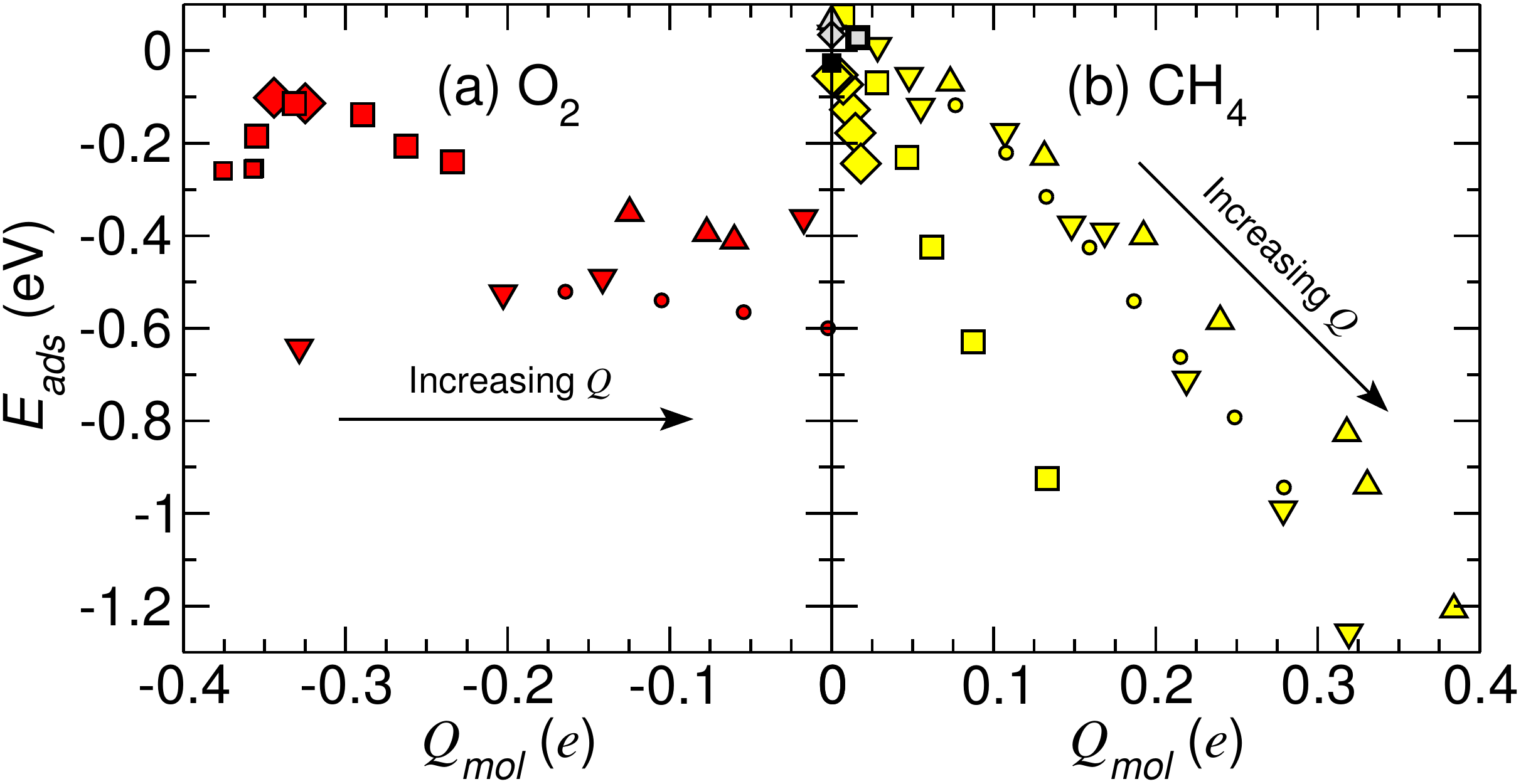}
\end{center}
\caption{ Adsorption energies $E_{\textit{ads}}$ in eV versus charge $Q_{mol}$ in $e$ of (a) O$_{\text{2}}$ and (b) CH$_{\text{4}}$ on Au$_{\text{2}}$ ($\circ$), Au$_{\text{6}}$ ($\triangle$), Au$_{\text{7}}$ ($\triangledown$), Au$_{\text{49}}$ ($\smallsquare$), Au$_{\text{55}}$ ($\square$), Au$_{\text{155}}$ ($\diamond$), and Au$_{\text{201}}$ ($\Diamond$) for increasing total charge $Q \geq $~0.  Grey (black) symbols denote adsorption on Au$_n$ (defects) on a rutile TiO$_{\text{2}}$(110) surface.
}\label{fgr:Fig4}
\noindent{\color{JPCCBlue}{\rule{\columnwidth}{1pt}}}
\end{figure}

To probe the influence of charging on the adsorption energies, we performed calculations with a fixed positive charge $Q$ applied to both the combined system and clean gold species.  Here the total charge $Q$ of an Au$_n$ species is in the range $0\leq Q \lesssim \frac{n}{\text{2}}e$.  For $Q \gtr \frac{n}{\text{2}}e$, Au$_n^Q$ begins to dissociate as the filling of all the \emph{s} levels is $ \lesssim \nicefrac{\text{1}}{\text{4}}$.  Figure~\ref{fgr:Fig4} shows the dependence of (a) O$_{\text{2}}$ and (b) CH$_{\text{4}}$ adsorption energies $E_{\textit{ads}}$ on the charge $Q_{mol}$ of the molecule adsorbed on the gold species.  The leftmost points in each panel of Figure~\ref{fgr:Fig4} correspond to the neutral system, with $Q$ increasing from left to right.  As the total charge $Q$ increases, the gold species becomes more electronegative, and the molecule's charge $Q_{mol}$ increases accordingly. 

For Au$_n$ on rutile TiO$_2$(110), we find CH$_4$ adsorbs on the corner sites of the gold nanoparticle.  We found CH$_4$ does not bind at the permimeter of the Au$_n$--TiO$_2$(110) interface.  We only found binding on the TiO$_2$(110) surface near an oxygen vacancy.  Although methane may bind at such defect sites, these sites do not appear to be catalytically active, as methane oxidation does not occur on the clean metal oxide surface.\cite{GuidoThesis}

On neutral Au$_n$ ($Q = $~0), the adsorption of O$_{\text{2}}$ is always stronger than that of CH$_{\text{4}}$ ($E_{\textit{ads}}[\text{O}_{\text{2}}] \ll E_{\textit{ads}}[\text{CH}_{\text{4}}]$). This is not surprising, as the adsorption of CH$_{\text{4}}$ to neutral Au$_n$ is negligible for $n > 2$, with $E_{\textit{ads}}[\text{CH}_{\text{4}}] \approx \text{-0.11}$~eV on Au$_{\text{2}}$.   We may relate this to the greater charging of O$_{\text{2}}$ ($-0.4\ e \lesssim Q_{mol} \lesssim -0.1\ e$) compared to CH$_{\text{4}}$ (0 $\lesssim Q_{mol} \lesssim \text{0.1}\ e$).  This means O$_{\text{2}}$ is more electronegative than any of the neutral gold clusters, which are in turn only slightly more electronegative than CH$_{\text{4}}$.  As the total charge $Q$ of the system increases, it becomes increasingly more difficult for O$_{\text{2}}$ to extract negative charge from Au$_n^Q$, increasing $Q_{mol}$.  This results in a weak dependence of $E_{\textit{ads}}[\text{O}_{\text{2}}]$ on $Q_{mol}$, compared to other factors such as cluster size and type. 
On the other hand, as $Q$ increases, it becomes easier for Au$_n^Q$ to extract negative charge from CH$_{\text{4}}$.  This provides the strong linear dependence of $E_{\textit{ads}}[\text{CH}_{\text{4}}]$ on $Q_{mol}$ shown in Figure~\ref{fgr:Fig4}(b).  In fact, upon sufficient charging $Q$ of Au$_n^Q$, we may tune the adsorption energy of CH$_{\text{4}}$ to be stronger than that of O$_{\text{2}}$.  For example, $E_{\textit{ads}}[\text{CH}_{\text{4}}] \approx -\text{0.42}$~eV and $E_{\textit{ads}}[\text{O}_{\text{2}}] \approx -\text{0.24}$~eV on Au$_{\text{55}}^{+4}$.  Under such conditions, with three orders of magnitude more CH$_{\text{4}}$ than O$_{\text{2}}$ adsorbed, the partial oxidation of CH$_{\text{4}}$ should be favoured on Au$_{\text{55}}^{+4}$.  

  In heterogeneous catalysis, each reaction barrier in a pathway is proportional to the sum of the reactants' adsorption energies \cite{HydrogenEvolution,FalsigCO,MowbrayJPCC, NorskovMethaneDecomposition}.  We have verified this for the first step in methane partial oxidation, which is methane decomposition ($\text{CH}_{\text{4}}\rightarrow \text{CH}_{\text{3}}+\text{H}$) on Au$_{\text{2}}^Q$ for $Q \geq 0$ \cite{LandmanCH4Au2}, as shown in the Supporting Information.  We also find the methane decomposition barrier is proportional to the total charge $Q$.
Thus, in the low coverage limit, the reactants' adsorption energies form a complete set of independent variables for determining the catalytic activity of gold nanoparticles.  In the case of methane partial oxidation, we see that $E_{\textit{ads}}[\text{CH}_{\text{4}}]$ is dependent on $Q$, while $E_{\textit{ads}}[\text{O}_{\text{2}}]$ is not.  This means we can use charging to tune the catalytic activity of gold nanoparticles for partial oxidation of methane under mild conditions.

Although the adsorption energy of CH$_{\text{4}}$ on Au$_n^Q$ is always proportional to the molecule's charge, i.e.~$E_{\textit{ads}}[\textrm{CH}_{\text{4}}] \approx E_0 - k Q_{mol}$, the constant of proportionality $k$ is dependent on the size and type of cluster.  Specifically, as the cluster increases in size, $k$ increases as well.  This is apparent from comparing the relative slopes of the adsorption relations shown in Figure~\ref{fgr:Fig4} for the different gold species. For example, for Au$_{\text{201}}^{+6}$, $E_{\textit{ads}}[\text{CH}_{\text{4}}] \approx -\text{0.24}$~eV and $Q_{mol} \approx \text{0.02}\ e$, with an Au--H separation of 2.7~\AA.  To understand the origin of these linear correlations and the differences in their slopes, we next analyze the electronic structure of these systems.

\begin{figure}
\noindent{\color{JPCCBlue}{\rule{\columnwidth}{1pt}}}
\begin{center}
\includegraphics[width=\columnwidth]{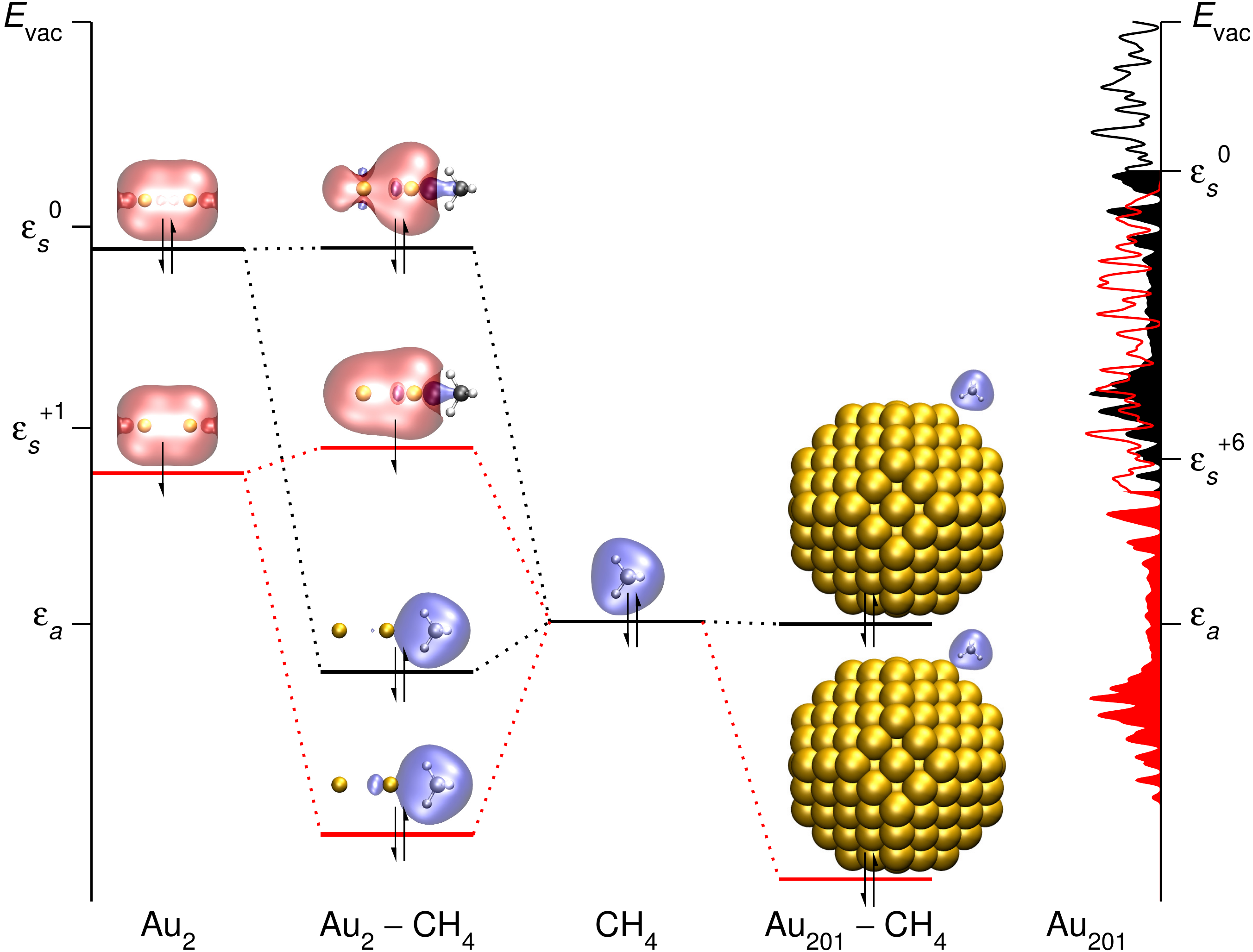}
\end{center}
\caption{ Schematic of the alignment and interaction between the CH$_{\text{4}}$ 2$a_{\text{1}}$ level at $\varepsilon_a$ and the 6\emph{s} levels centered at $\varepsilon_s$ of: (left) Au$_{\text{2}}$ (black), Au$_{\text{2}}^{\text{+1}}$ (red); (right) Au$_{\text{201}}$ (black), and Au$_{\text{201}}^{+6}$ (red).  Filling denotes occupancy.
}\label{fgr:Fig5}
\noindent{\color{JPCCBlue}{\rule{\columnwidth}{1pt}}}
\end{figure}

Figure~\ref{fgr:Fig5} shows the electronic interaction between CH$_{\text{4}}$ and the smallest (Au$_{\text{2}}$) and largest (Au$_{\text{201}}$) gold species considered.  In both cases we may describe the adsorption of CH$_{\text{4}}$ through the occupied 6\emph{s} levels of gold and the 2$a_{\text{1}}$
 orbital of methane, as shown in Figure~\ref{fgr:Fig5}.  We find the charge transfer $Q_{mol}$ from CH$_{\text{4}}$ to Au$_n$ is through level hybridization, and not through occupation of the Au$_n$ unoccupied levels.

In particular, for Au$_{\text{2}}$ there is strong hybridization of the $6s$ and 2$a_{\text{1}}$ levels into the bonding and anti-bonding orbitals shown in Figure~\ref{fgr:Fig5}.  When we positively charge Au$_{\text{2}}$, we find more weight of the 2$a_{\text{1}}$--6\emph{s} bonding orbital on the binding Au atom, and less weight of the 2$a_{\text{1}}$--6\emph{s} anti-bonding orbital on methane as it is emptied.

On the other hand, the interaction of CH$_{\text{4}}$ with Au$_{\text{201}}$ is through a distribution of many 6\emph{s} levels, which are described by the projected density of states (PDOS) onto the gold 6\emph{s} levels, $\rho_s(\varepsilon)$.  However, the interaction is mainly described by the average energy and bandwidth of $\rho_s(\varepsilon)$.
Figure~\ref{fgr:Fig5} shows the 2$a_1$--$6s$ totally bonding orbital between methane and Au$_{\text{201}}^{\text{0}}$ or Au$_{\text{201}}^{+\text{6}}$.  
  The Au$_{\text{201}}^{\text{0}}$ 6\emph{s} levels are all above the methane 2$a_{\text{1}}$ level, which is unchanged upon adsorption. However, the Au$_{\text{201}}^{+6}$ 6\emph{s} levels are aligned with the methane 2$a_{\text{1}}$ level, which is strongly renormalized upon adsorption.

The Au$_n$ -- CH$_{\text{4}}$ interaction is primarily through the 6\emph{s} levels because the 5\emph{d} levels of gold are always completely filled.  This means the corresponding bonding and anti-bonding levels with methane are also filled.  Consequently, the 5\emph{d} levels of gold only contribute a weak repulsive interaction to the adsorption \cite{Norskovd-bandmodelSurfSci1995}.  

In fact, positively charging Au$_n^Q$ ($Q \lesssim \frac{n}{2}e$) only changes the filling of the bonding 6\emph{s} levels of gold.  This is shown for Au$_{\text{2}}^{\text{+1}}$ in Figure~\ref{fgr:Fig5}.  It is these changes in filling that result in a half-empty anti-bonding $6s$--2$a_{\text{1}}$ level for Au$_{\text{2}}^{\text{+1}}$ -- CH$_{\text{4}}$.  We shall show that CH$_{\text{4}}$ adsorption is via the emptying of the anti-bonding 2$a_{\text{1}}$--6\emph{s} level.

Overall, charging rigidly shifts all the gold levels down in energy relative to the vacuum energy $E_{\textit{vac}}$.   This brings the $6s$ levels of gold into better alignment with the occupied 2$a_{\text{1}}$ level of methane $\varepsilon_a$, resulting in stronger hybridization and adsorption.  

To quantify this shift, we use the average energy of the 6\emph{s} levels of the gold species, $\varepsilon_s \equiv \int \varepsilon \rho_{s}(\varepsilon) d\varepsilon$.   This quantity is analogous to the \emph{d}-band center commonly used as a simple descriptor for bulk and surface systems. 
As shown in Figure~\ref{fgr:Fig5}, \(\varepsilon_s^Q - \varepsilon_a \ll \varepsilon_s^{\text{0}} - \varepsilon_a\) for both Au$_{\text{2}}^{\text{+1}}$ and Au$_{\text{201}}^{+6}$.  This results in a significantly stronger adsorption of CH$_{\text{4}}$ on the charged gold species, as shown in Figure~\ref{fgr:Fig4}.  

\begin{figure}[!thb]
\begin{center}
\includegraphics[width=0.99\columnwidth]{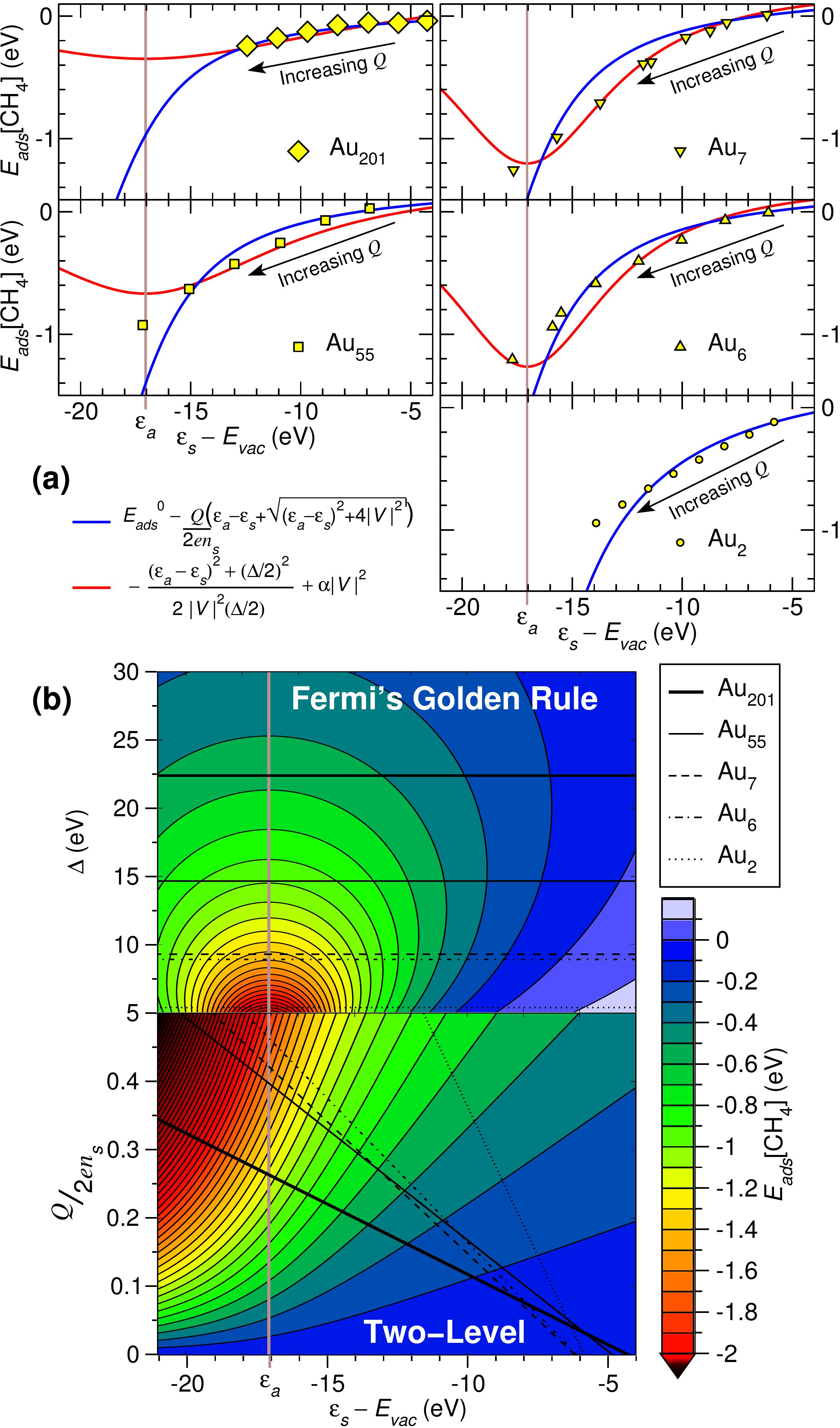}
\end{center}
\caption{(a) Methane adsorption energy $E_{\textit{ads}}[\text{CH}_{\text{4}}]$ in eV versus average energy of the gold 6\emph{s} levels $\varepsilon_s$ in eV relative to the vacuum level $E_{\textit{vac}}$ for Au$_{\text{2}}$ ($\circ$), Au$_{\text{6}}$ ($\triangle$), Au$_{\text{7}}$ ($\triangledown$), Au$_{\text{55}}$ ($\square$), and Au$_{\text{201}}$ ($\Diamond$) for increasing total charge $Q \geq $~0.  A Fermi's Golden Rule based fit (red) and a two-level model fit (blue) are provided for comparison.
(b) Contour plot of methane adsorption energy $E_{\textit{ads}}[\text{CH}_{\text{4}}]$ in eV from (top) Fermi's Golden Rule model versus width of the gold 6\emph{s} levels $\Delta$ in eV and (bottom) two-level model versus emptying of the gold 6\emph{s} levels $\frac{Q}{\text{2}en_s}$ and the average energy of the 6\emph{s} levels $\varepsilon_s$ in eV relative to the vacuum level $E_{\textit{vac}}$ for Au$_{\text{2}}$, Au$_{\text{6}}$, Au$_{\text{7}}$, Au$_{\text{55}}$, and Au$_{\text{201}}$.
}\label{fgr:Fig6}
\noindent{\color{JPCCBlue}{\rule{\columnwidth}{1pt}}}
\end{figure}

Based on this analysis, we use $\varepsilon_s$ as a descriptor for Au$_n$ -- CH$_{\text{4}}$ adsorption.  Figure~\ref{fgr:Fig6}(a) shows the dependence of the methane adsorption energy on $\varepsilon_s$ for Au$_n$, with $n = $ 2, 6, 7, 55, 201.  As Au$_n$ is charged, the 6\emph{s} levels are shifted down in energy, i.e.\ $\varepsilon_s$ decreases relative to $E_{\textit{vac}}$, and CH$_{\text{4}}$ is more strongly bound.

The methane adsorption may be modelled semi-quantitatively including only the interaction between the 2$a_{\text{1}}$ level of CH$_{\text{4}}$ at $\varepsilon_a$ and a typical 6\emph{s} bonding orbital of Au$_n^Q$ at $\varepsilon_s$ with an occupancy of $2 - \frac{Q}{en_s}$ electrons.  Here $n_s$ is the number of \emph{s} levels which are available for bonding.  Solving this simple two-level model, we find the adsorption energy for methane is given by
\begin{equation}
E_{\textit{ads}}\ \approx\ E_{\textit{ads}}^{0} - \frac{Q}{2en_s}\left(\varepsilon_a - \varepsilon_s + \sqrt{(\varepsilon_a-\varepsilon_s)^2 + 4|V|^2}\right)\label{eqn:2levelanswer}
\end{equation}
\begin{equation}
\ \ \ \approx\ E_{\textit{ads}}^{0} - \frac{Q}{2en_s}\frac{2|V|^2}{|\varepsilon_a - \varepsilon_s|}, \textrm{ for } |V| \ll |\varepsilon_a - \varepsilon_s|,\label{eqn:2levelVsmall}
\end{equation}
where $E_{\textit{ads}}^{0}$ is the adsorption energy on the neutral gold species and $|V|^2 \approx \text{3.40}$~eV$^{\text{2}}$ is the coupling matrix element between the 6\emph{s} levels of Au$_n$ and the 2$a_{\text{1}}$ level of CH$_{\text{4}}$.  Such a simple model has previously proven effective for describing the binding with the \emph{d}-band of H$_{\text{2}}$ on transition metal surfaces \cite{Norskovd-bandmodelSurfSci1995}. A derivation of the two-level model and the parameters employed is provided in Supporting Information.

We find that this simple two-level model describes the adsorption of CH$_{\text{4}}$ on Au$_n$ semi-quantitatively, as shown in Figure~\ref{fgr:Fig6}(a).  Here we have assumed all the 6\emph{s} levels play a role in the adsorption for the small clusters ($n = $ 2, 6, and 7), while for the larger gold nanoparticles we have included 15 and 37 6\emph{s} levels for Au$_{\text{55}}$ and Au$_{\text{201}}$, respectively.  However, for the larger gold nanoparticles the methane 2$a_{\text{1}}$ level interacts with a distribution of many gold 6\emph{s} levels, i.e.\ $\rho_s(\varepsilon)$.  In this case, the adsorption energy is $-\hslash$ times the total transition rate $W_{a\rightarrow s}$ between the methane 2$a_{\text{1}}$ level and the effective continuum of gold 6\emph{s} levels.  The total transition rate may be obtained to lowest order from Fermi's Golden Rule, so that
\begin{equation}
E_{\textit{ads}} \approx -\hslash W_{a\rightarrow s} \approx - 2 \pi |V|^2 \widetilde{\rho}_s(\varepsilon_a)\approx - \frac{2|V|^2(\Delta/2)}{(\varepsilon_a - \varepsilon_s)^2 + (\Delta/2)^2},\label{eqn:fmol}
\end{equation}
where $\widetilde{\rho}_s(\varepsilon)$ is the effective density of 6\emph{s} levels of Au$_n^Q$.  In the wide-band limit, $\widetilde{\rho}_s(\varepsilon)$ is a Lorentzian distribution centered at $\varepsilon_s$ with a lifetime $\tau_s \approx \hslash/\Delta$, where $\Delta$ is the width of the 6\emph{s} band of Au$_n^Q$.  As shown in Figure~\ref{fgr:Fig6}, the 2$a_{\text{1}}$ level is at \(\varepsilon_a\approx -\text{17.05}\)~eV.  Note that Equation~\eqref{eqn:fmol} neglects the small repulsive interaction from the 5\emph{d} levels of gold.  This may be incorporated through an additional $\alpha |V|^2$ term \cite{Norskovd-bandmodelSurfSci1995}, where $\alpha \approx \text{0.07}$~eV$^{\text{-1}}$.  

From Figure~\ref{fgr:Fig6}(a) we see that such a Fermi's Golden Rule based fit proves quite effective in describing the adsorption of CH$_{\text{4}}$ to the Au$_n$ clusters for $n=$ 6, 7, 55, 201.  It should be noted that this procedure incorporates only two fitting parameters: the Au -- CH$_{\text{4}}$ coupling matrix elements $V$ and the coupling constant $\alpha$.  Since $V$ and $\alpha$ depend only on the coupling between CH$_{\text{4}}$ and the metal \emph{s} levels, they are expected to have similar values for all the gold species considered.  This is indeed the case.  On the other hand, the Fermi's Golden Rule based model breaks down for Au$_{\text{2}}$.  This is not surprising, as $\rho_s$ for Au$_{\text{2}}$ is a strongly bimodal distribution, which cannot be described by a Lorentzian.  A derivation of the Fermi's Golden Rule model and the parameters employed is provided in Supporting Information.

Figure~\ref{fgr:Fig6}(b) shows how the methane adsorption energy $E_{\textit{ads}}$ depends on the average energy of the \emph{s} levels $\varepsilon_s$, the bandwidth $\Delta$ in the Fermi's Golden Rule model, and the emptying of the metal \emph{s} level $\frac{Q}{2en_s}$ in the two-level model. As the nanoparticle's size increases, so does the bandwidth, resulting in a weaker methane adsorption according to the Fermi's Golden Rule model. In the two-level model, emptying the metal \emph{s} level by the same amount shifts $\varepsilon_s$ down by a smaller amount for smaller nanoparticles.  Because of this, smaller nanoparticles require a greater emptying of the \emph{s} level to obtain the same methane adsorption.
Using Figure~\ref{fgr:Fig6}(b), one may estimate the adsorption energy of methane on nanoparticles of arbitrary size, postitive charge, and metallic composition based on only the average energy of the \emph{s} levels and either their bandwidth or emptying.  In this way, one may use charging to tune $\varepsilon_s$, and thus control and predict methane's adsorption energy.  This is a critically important first step towards tuning the catalytic activity of metal nanoparticles.

\begin{figure}
\noindent{\color{JPCCBlue}{\rule{\columnwidth}{1pt}}}
\begin{center}
\includegraphics[width=0.74\columnwidth]{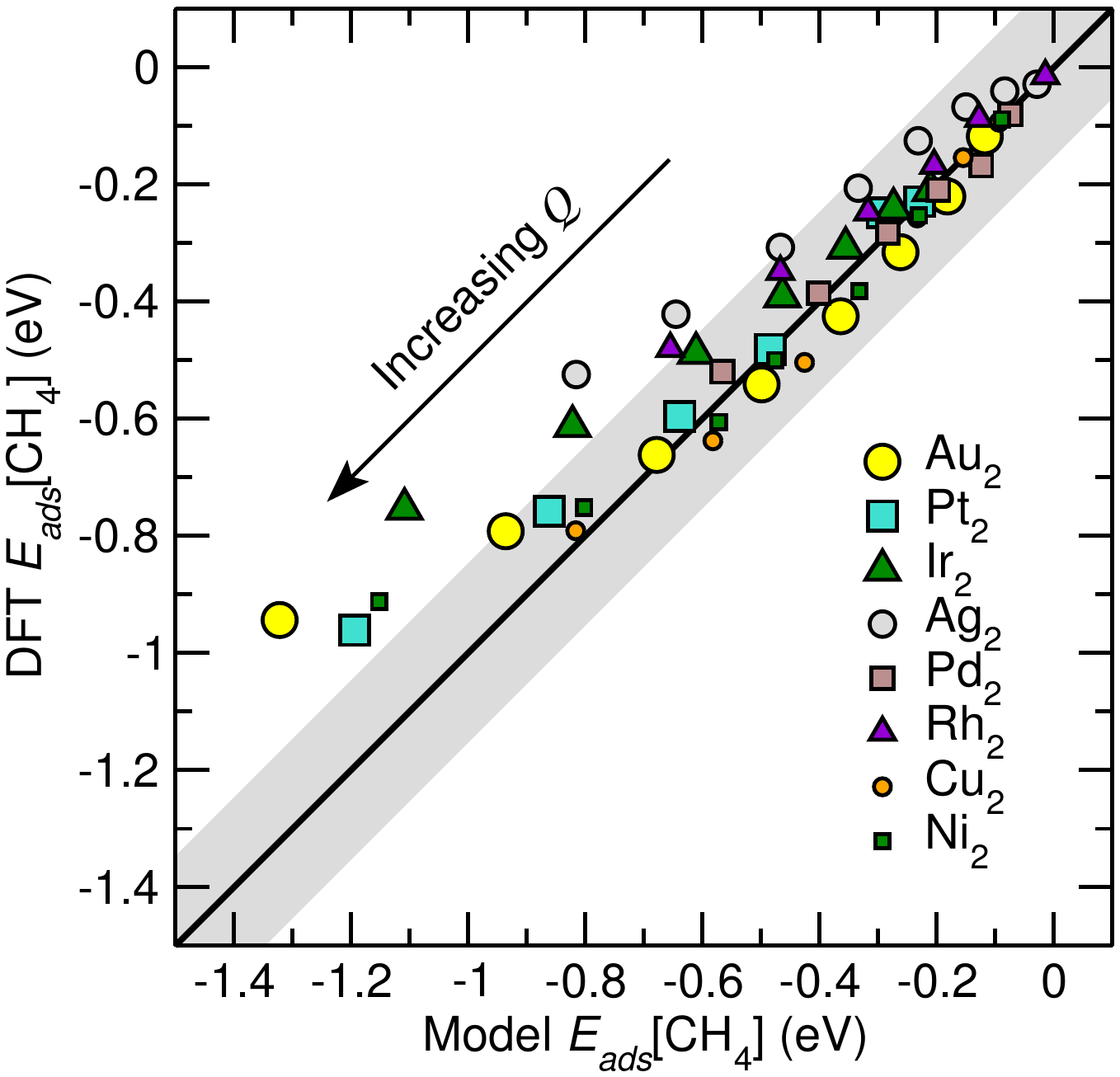}
\end{center}
\caption{ Methane adsorption energy $E_{\textit{ads}}[\text{CH}_{\text{4}}]$ in eV from DFT versus a two-level model for Au$_{\text{2}}$ ($\largecircle$), Pt$_{\text{2}}$ ($\largesquare$), Ir$_{\text{2}}$ ($\largetriangleup$), Ag$_{\text{2}}$ ($\medcirc$), Pd$_{\text{2}}$ ($\square$), Rh$_{\text{2}}$ ($\triangle$), Cu$_{\text{2}}$ ($\circ$), and Ni$_{\text{2}}$ ($\smallsquare$) for increasing total charge $Q \geq $~0.  Grey regions depict the standard deviation.
}\label{fgr:Fig7}
\noindent{\color{JPCCBlue}{\rule{\columnwidth}{1pt}}}
\end{figure}

In Figure~\ref{fgr:Fig7} we compare the two-level model with DFT for CH$_{\text{4}}$ adsorption on all fcc transition metal dimers (X$_{\text{2}}$ for X $\in \{$Au, Pt, Ir, Ag, Pd, Rh, Cu, Ni$\}$).  We find $E_{\textit{ads}}[\text{CH}_{\text{4}}]$ increases in strength as the metal dimers are more positively charged ($Q \geq $~0).  The two-level model describes the calculated adsorption energies for CH$_{\text{4}}$ on X$_{\text{2}}$ semi-quantitatively, with a standard deviation of $\sigma \approx \pm$0.15~eV.  This confirms that the interaction with the metal dimers is mostly between the occupied metal \emph{s} levels and the methane 2$a_{\text{1}}$ level.  

We have shown that by positively charging gold nanoparticles, one may tune the adsorption of CH$_{\text{4}}$.  This is in contrast to O$_{\text{2}}$ adsorption. This suggests through surface or electrostatic doping it may be possible to control the charge of the gold cluster, thereby making CH$_{\text{4}}$ adsorb stronger than O$_{\text{2}}$, and allowing one to perform delicate oxidation on gold.  Further, the success of the predictive models employed suggests the mechanism of CH$_{\text{4}}$ adsorption is predominantly hybridization between the 2$a_{\text{1}}$ level of methane and the \emph{s} levels of the metal.  This result is general, applying from small clusters to large nanoparticles, and for all fcc transition metals dimers.  The mechanism we propose provides a new strategy to tune both the adsorption and catalytic activities of organic molecules on metallic nanoparticles.

\titleformat{\section}{\bfseries\sffamily\color{JPCCBlue}}{\thesection.~}{0pt}{\large$\blacksquare$\normalsize~}

\section*{ASSOCIATED CONTENT}
\subsubsection*{\includegraphics[height=8pt]{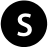} Supporting Information}
\noindent Comparison of xc-functionals, methane decomposition barrier, model derivations, model parameters, total energies, optimized geometries, and oxygen reference energy.  This material is available free of charge via the Internet at http://pubs.acs.org.

\section*{AUTHOR INFORMATION}
\subsubsection*{Corresponding Authors}
\noindent *E-mail: duncan.mowbray@gmail.com (D.J.M.).\\ *E-mail: annapaola.migani@cin2.es (A.M.).\\ *E-mail: angel.rubio@ehu.es (A.R.).
\subsubsection*{Notes} 
\noindent The authors declare no competing financial interest.
\section*{ACKNOWLEDGMENTS} 
We acknowledge funding by the European Research Council Advanced Grant DYNamo (ERC-2010-AdG-267374), Spanish Grants (FIS2010-21282-C02-01 and PIB2010US-00652), Grupos Consolidados UPV/EHU del Gobierno Vasco (IT-578-13), and the European Commission project CRONOS (280879-2 CRONOS CP-FP7).  DJM acknowledges funding through the Spanish ``Juan de la Cierva'' program (JCI-2010-08156).  AM acknowledges funding through JAE DOC and the European Social Fund. GW acknowledges funding through the HPC-Europa2 Programme (Project No.\ 228398).

\bibliography{bibliography}

\end{document}